\documentclass[prc,twocolumn,floatfix,groupedaddress,nofootinbib,showpacs,preprintnumbers,
amsmath,amssymb,amsfonts,superscriptaddress,widetable] {revtex4}
\usepackage{bm}
\usepackage{mathrsfs}
\usepackage{amssymb}
\usepackage{amsmath}
\usepackage{graphicx}
\usepackage{array}
\usepackage{color}

\begin{document}
%%%%%%%%%%%
%%%%%%%%%%%
\title{Neutron skins and neutron stars in the multi-messenger era}
%%%
\author{F.~J. Fattoyev}\email{ffattoye@indiana.edu}
\affiliation{Center for Exploration of Energy and Matter and
                  Department of Physics, Indiana University,
                  Bloomington, IN 47405, USA}
\author{J. Piekarewicz}\email{jpiekarewicz@fsu.edu}
\affiliation{Department of Physics, Florida State University,
               Tallahassee, FL 32306, USA}
\author{C. J. Horowitz}\email{horowit@indiana.edu}
\affiliation{Center for Exploration of Energy and Matter and
                  Department of Physics, Indiana University,
                  Bloomington, IN 47405, USA}
%%%%%%%%%%%
\date{\today}
\begin{abstract}
The historical first detection of a binary neutron star merger by
the LIGO-Virgo collaboration [B. P. Abbott {\sl et al.} Phys. Rev.
Lett. 119, 161101 (2017)] is providing fundamental new insights into
the astrophysical site for the $r$-process and on the nature of
dense matter. A set of realistic models of the equation of state
(EOS) that yield an accurate description of the properties of finite
nuclei, support neutron stars of two solar masses, and provide a
Lorentz covariant extrapolation to dense matter are used to
confront its predictions against tidal polarizabilities extracted
from the gravitational-wave data. Given the sensitivity of the
gravitational-wave signal to the underlying EOS, limits on the tidal
polarizability inferred from the observation translate into
constraints on the neutron-star radius. Based on these
constraints, models that predict a stiff symmetry energy, and thus
large stellar radii, can be ruled out. Indeed, we deduce an upper
limit on the radius of a $1.4\,M_{\odot}$ neutron star of
$R_{\star}^{1.4}\!<\!13.76\,{\rm km}$. Given the sensitivity of the
neutron-skin thickness of ${}^{208}$Pb to the symmetry energy,
albeit at a lower density, we infer a corresponding upper limit of
about $R_{\rm skin}^{208}\!\lesssim\!0.25\,{\rm fm}$. However, if
the upcoming PREX-II experiment measures a significantly thicker
skin, this may be evidence of a softening of the symmetry energy
at high densities---likely indicative of a phase transition in the interior
of neutron stars.

\end{abstract}
\smallskip

\maketitle

\emph{What are the new states of matter at exceedingly high density
and temperature?} and \emph{how were the elements from iron to uranium
made?} are two of the ``eleven science questions for the next century"
identified by the National Academies Committee on the Physics of the
Universe\,\cite{QuarksCosmos:2003}. In framing these questions, the
committee recognized the deep connections between the very small
and the very large. In one clean sweep, the historical first detection of
a binary neutron star (BNS) merger by the LIGO-Virgo
collaboration\,\cite{Abbott:PRL2017} has started to answer these
fundamental questions by providing critical insights into the nature of
dense matter and on the synthesis of the heavy elements.

Gravitational waves (GW) from the BNS merger GW170817 emitted from a
distance of about 40 Mpc were detected by the LIGO
gravitational-wave observatory\,\cite{Abbott:PRL2017}. About two
seconds later, the Fermi Gamma-ray Space Telescope
(Fermi)\,\cite{Goldstein:2017mmi} and the International Gamma-Ray
Astrophysics Laboratory (INTEGRAL)\,\cite{Savchenko:2017ffs}
identified a short duration $\gamma$-ray burst associated with the
BNS merger. Within eleven hours of the GW detection, ground- and
spaced-based telescopes operating at a variety of wavelengths
identified the associated \emph{kilonova}---the electromagnetic
transient powered by the radioactive decay of the heavy elements
synthesized in the rapid neutron-capture process ($r$-process).
Characteristic features of the optical spectrum are consistent with
the large opacity typical of the lanthanides (atomic number 57--71)
and have revealed that about 0.05 solar masses (or about $10^{4}$
earth masses) of $r$-process elements were synthesized in this
single event\,\cite{Cowperthwaite:2017dyu, Chornock:2017sdf,
Nicholl:2017ahq}. The gravitational wave detection from the BNS
merger, together with its associated electromagnetic counterparts,
open the new era of \emph{multi-messenger} astronomy and provide
compelling evidence in favor of the long-held belief that
neutron-star mergers play a critical role in the production of heavy
elements in the cosmos.

Besides the identification of the BNS merger as a dominant site for
the $r$-process, such an unprecedented event imposes significant
constraints on the EOS of dense matter. In particular, the
\emph{tidal polarizability} (or deformability) is an intrinsic
neutron-star property highly sensitive to the stellar
compactness\,\cite{Hinderer:2007mb,Hinderer:2009ca,Damour:2009vw,
Postnikov:2010yn,Fattoyev:2012uu,Steiner:2014pda} that describes the
tendency of a neutron star to develop a mass quadrupole as a
response to the tidal field induced by its
companion\,\cite{Damour:1991yw,Flanagan:2007ix}. The dimensionless
tidal polarizability $\Lambda$ is defined as follows:
%%%
\begin{equation}
 \Lambda = \frac{2}{3}k_{2}\left(\frac{c^{2}R}{GM}\right)^{5}
                 =\frac{64}{3}k_{2}\left(\frac{R}{R_{s}}\right)^{5}\;,
 \label{Lambda}
\end{equation}
%%%
where $k_{2}$ is the second Love number\,\cite{Binnington:2009bb,
Damour:2012yf}, $M$ and $R$ are the neutron star mass and radius,
respectively, and $R_{s}\!\equiv\!2GM/c^{2}$ is the Schwarzschild
radius. A great virtue of the tidal polarizability is its high
sensitivity to the stellar radius ($\Lambda\!\sim\!\!R^{5}$) a
quantity that has been notoriously difficult to
constrain\,\cite{Ozel:2010fw,Steiner:2010fz,Suleimanov:2010th,
Guillot:2013wu,Lattimer:2013hma,Heinke:2014xaa,Guillot:2014lla,
Ozel:2015fia,Watts:2016uzu,Steiner:2017vmg,Nattila:2017wtj}.
Pictorially, a ``fluffy" neutron star having a large radius is much
easier to polarize than the corresponding compact star with the same
mass but a smaller radius. Finally, a derived quantity from the
individual tidal polarizabilities $\Lambda_{1}$ and $\Lambda_{2}$
related to the phase of the gravitational
wave\,\cite{Flanagan:2007ix, Damour:2009wj,
Baiotti:2010xh,Damour:2012yf} is given by
%%%
\begin{equation}
 \widetilde{\Lambda} \!=\! \frac{16}{13} \!\left[
   \frac{(M_{1}\!+\!12M_{2})M_{1}^{4}}{(M_{1}\!+\!M_{2})^{5}}\Lambda_{1} \!+\!
   \frac{(M_{2}\!+\!12M_{1})M_{2}^{4}}{(M_{1}\!+\!M_{2})^{5}}\Lambda_{2} \right].
   \label{LambdaTilde}
\end{equation}
%%%
Note that for the equal-mass case,
$\widetilde{\Lambda}\!=\!\Lambda_{1}\!=\!\Lambda_{2}$. Remarkably,
the tidal polarizability determined from the first BNS merger is already
stringent enough to rule out a significant number of previously viable
EOSs\,\cite{Abbott:PRL2017}.

In this letter we explore in greater detail the impact of the BNS merger
on the EOS and on those laboratory observables that are particularly
sensitive to the nuclear symmetry energy---a quantity that represents
the increase in the energy of the system as it departs from the symmetric
limit of equal number of neutrons and protons; see
Refs.\,\cite{Tsang:2012se,Li:2014,Horowitz:2014bja} and references
contained therein. Particularly uncertain is the density dependence of the
symmetry energy, often encoded in a quantity denoted by $L$ that is closely
related to the pressure of pure neutron matter at saturation density.
%($\rho_{0}\!\simeq\!0.15\,{\rm fm}^{-3}$).

A laboratory observable that has been identified as strongly correlated to
both $L$ and to the radius of low-mass neutron stars is the
\emph{neutron-skin thickness} of atomic nuclei---defined as the difference
between the neutron ($R_{n}$) and proton ($R_{p}$) root-mean-square radii:
$R_{\rm skin}\!=\!R_{n}\!-\!R_{p}$. Despite a difference in length scales of
19 orders of magnitude, the size of a neutron star and the thickness
of the neutron skin share a common origin: the pressure of
neutron-rich matter. That is, whether pushing against surface
tension in an atomic nucleus or against gravity in a neutron star,
both the neutron skin and the stellar radius are sensitive to the
same EOS.

The pioneering Lead Radius Experiment (PREX) at the
Jefferson Laboratory has provided the first model-independent evidence
in favor of a neutron-rich skin in
${}^{208}$Pb\,\cite{Abrahamyan:2012gp,Horowitz:2012tj}:
$R_{\rm skin}^{208}\!=\!{0.33}^{+0.16}_{-0.18}\,{\rm fm}.$
%%%
%\begin{equation}
%   R_{\rm skin}^{208}\!\equiv\!
%   R_{n}({}^{208}{\rm Pb}) - R_{p}({}^{208}{\rm Pb}) =
%   {0.33}^{+0.16}_{-0.18}\,{\rm fm}.
%  \label{PREX}
%\end{equation}
%%%
Although the central value is significantly larger than suggested by
most theoretical predictions, the large statistically-dominated
uncertainty prevents any real tension between theory and experiment.
In an effort to impose meaningful theoretical constraints, an approved
follow-up experiment (PREX-II) is envisioned to reach a $0.06$\,fm
sensitivity.

%%%%%%%%%%%%%%%%%%%%%%%%%%%%%%%%%%%%
% Updated; Starting with Jorge's Original Writings
%%%%%%%%%%%%%%%%%%%%%%%%%%%%%%%%%%%%
To connect the tidal polarizability to nuclear observables sensitive to the
density dependence of the symmetry energy\,\cite{Fattoyev:2012uu}, we
model the EOS using a relativistic mean-field (RMF) approach pioneered
by Serot and Walecka\,\cite{Walecka:1974qa,Serot:1984ey} which has
been continuously improved throughout the years\,\cite{Boguta:1977xi,
Mueller:1996pm,Horowitz:2000xj,Todd-Rutel:2005fa}. The effective
Lagrangian density is written exclusively in terms of conventional degrees
of freedom (neutrons, protons, electrons, and muons) and includes a handful
of parameters that are calibrated to provide an accurate description of finite
nuclei and---critically to the description of neutron stars---a \emph{Lorentz
covariant} extrapolation to dense nuclear matter. Although increasingly
sophisticated fitting protocols are now able to incorporate more stringent
constraints from
finite nuclei and neutron stars\,\cite{Chen:2014sca}, the \emph{isovector}
sector of the effective Lagrangian---responsible for generating the density
dependence of the symmetry energy---remains largely unconstrained. To
mitigate this problem we follow a simple procedure first proposed in
Ref.\,\cite{Horowitz:2000xj} that enables one to fine tune the value of the
slope of the symmetry energy $L$ without compromising the success of
the model in reproducing well measured observables. We label the set of
models generated in this manner the ``FSUGold2 family".

%%%%%%%%%%%%%%%%%%%%%%%%%%%%%%%%%%%%
\begin{figure}[ht]
\smallskip
 \includegraphics[width=0.9\columnwidth]{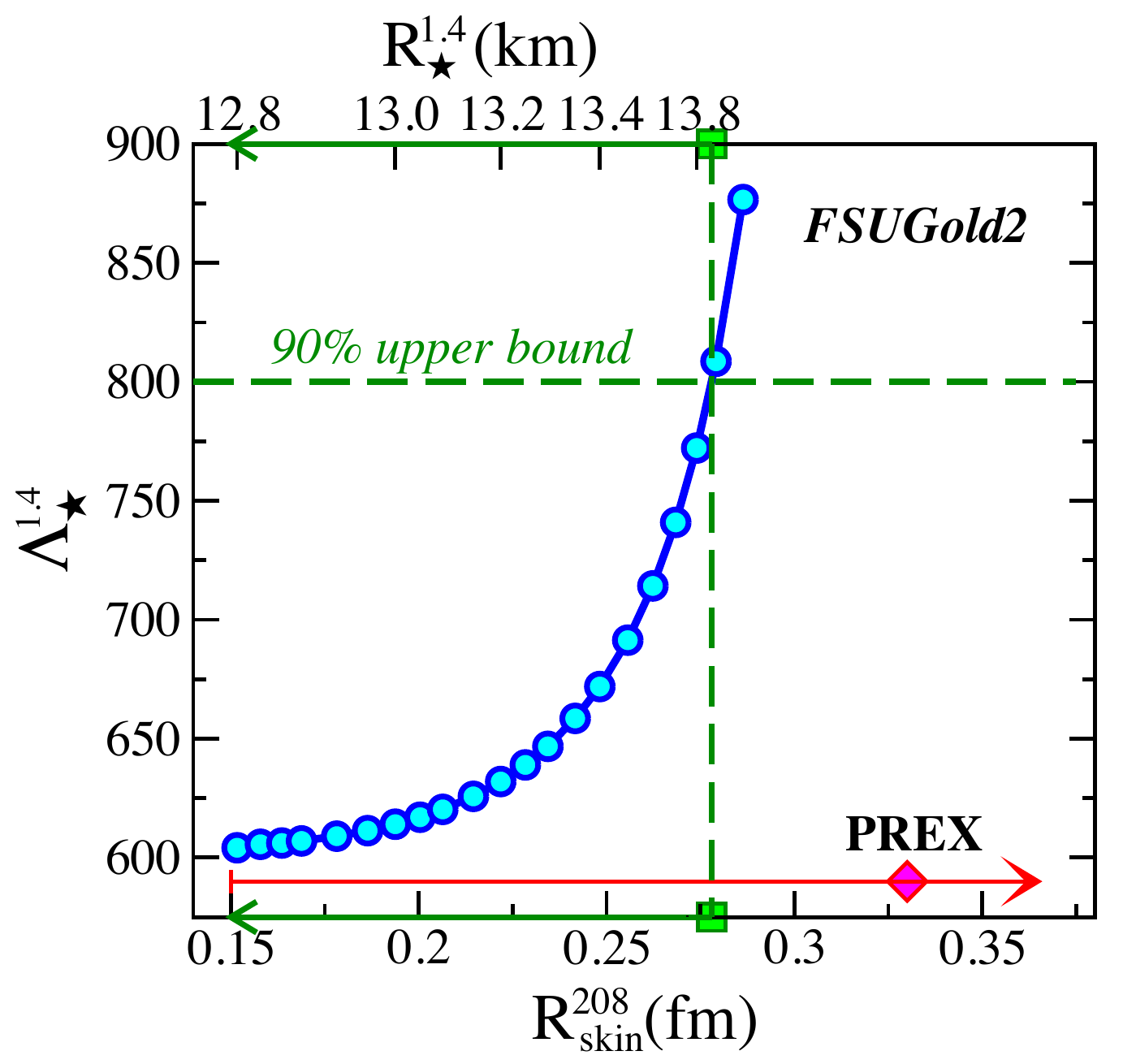}
 \caption{(Color online). The dimensionless tidal
 polarizability $\Lambda^{1.4}_{\star}$ of a $1.4\,M_{\odot}$
 neutron star as a function of the neutron-skin thickness of
 ${}^{208}$Pb (lower abscissa) and the radius of a
 $1.4\,M_{\odot}$  neutron star (upper abscissa) as predicted
 by the FSUGold2 family of relativistic interactions. Constraints
 on $R_{\rm skin}^{208}$ and $R_{\star}^{1.4}$ are inferred from
 adopting the $\Lambda^{1.4}_{\star}\!\le\!800$ limit deduced
 from GW170817\,\cite{Abbott:PRL2017}.}
 \label{Fig1}
\end{figure}
%%%%%%%%%%%%%%%%%%%%%%%%%%%%%%%%%%%%

In Fig.\,\ref{Fig1} we use the FSUGold2 family to predict the
tidal polarizability $\Lambda^{1.4}_{\star}$ of a $1.4\,M_{\odot}$ neutron star
as a function of both $R_{\rm skin}^{208}$ and $R_{\star}^{1.4}$
(the radius of a $1.4\,M_{\odot}$ neutron star). It is important to underscore
that the predictions for all three observables displayed in the figure are generated
from the same interaction. That is, for each member of the FSUGold2 family,
the model parameters remain unchanged in going from finite nuclei to neutron
stars.
%Results are shown for the equal
%mass scenario of $M_{1}\!=\!M_{2}\!=\!1.365\,M_{\odot}$ and for the
%unequal case of $M_{1}\!=\!1.60\,M_{\odot}$ and
%$M_{2}\!=\!1.17\,M_{\odot}$---both corresponding to the very well
%constrained chirp mass of ${\cal M}\!=\!(M_{1}M_{2})^{3/5}
%(M_{1}\!+\!M_{2})^{-1/5}\!\!=\!1.188\,M_{\odot}$\,\cite{Abbott:PRL2017}.
As anticipated, the 90\% confidence limit on $\Lambda^{1.4}_{\star}\!\le\!800$
extracted from the GW signal translates into a corresponding upper limit
on the radius of a $1.40\,M_{\odot}$ neutron star of
$R_{\star}^{1.4}\!\le\!13.9\,{\rm km}$. Also shown in the figure is the central
value of $R_{\rm skin}^{208}$ as measured by the PREX
collaboration\,\cite{Abrahamyan:2012gp,Horowitz:2012tj}, with the
red arrow highlighting the rather large experimental uncertainty.
Adopting the $\Lambda^{1.4}_{\star}\!\le\!800$ limit excludes the
$R_{\rm skin}^{208}\!\gtrsim\!0.28$\,fm region---suggesting that
the neutron-skin thickness of ${}^{208}$Pb cannot be overly large.
However, if the large value of $R_{\rm skin}^{208}$ is confirmed by
PREX-II, then an intriguing scenario may develop. A thick neutron
skin would suggest that the EOS at the typical densities found in
atomic nuclei is stiff, while the small neutron-star radii inferred
from the BNS merger implies that the EOS at higher densities is
soft. The evolution from stiff to soft may be indicative of a phase
transition in the interior of neutron stars.

While the FSUGold2 family provides the flexibility to generate a continuum
of realistic models with varying neutron skins, the models span a fairly
narrow range of neutron-star radii (see Fig.\,\ref{Fig1}). To alleviate this
problem---and in the spirit of Ref.\,\cite{Abbott:PRL2017}---we provide
predictions using a representative set of RMF models. As in the case
of the FSUGold2 family, these models are successful in reproducing
laboratory observables and are also consistent with the
$M_{\star}\!=\!2.01\,\pm\,0.04\,M_{\odot}$ limit\,\cite{Demorest:2010bx,
Antoniadis:2013pzd}. Yet, being less restrictive than the FSUGold2
family, they can generate a wider range of stellar radii. For reference,
the ten models adopted in this letter are:
NL3\,\cite{Lalazissis:1996rd, Lalazissis:1999},
IU-FSU\,\cite{Fattoyev:2010mx}, %(the maximum mass constraint for this
%model is reproduced at a 2-$\sigma$ level),
TAMUC-FSU\,\cite{Fattoyev:2013yaa}, FSUGold2\,\cite{Chen:2014sca},
and FSUGarnet together with three parametrizations denoted by
RMF022, RMF028, and RMF032\,\cite{Chen:2014mza}.

%%%%%%%%%%%%%%%%%%%%%%%%%%%%%%%%%%%%
\begin{figure}[ht]
\smallskip
 \includegraphics[width=0.9\columnwidth]{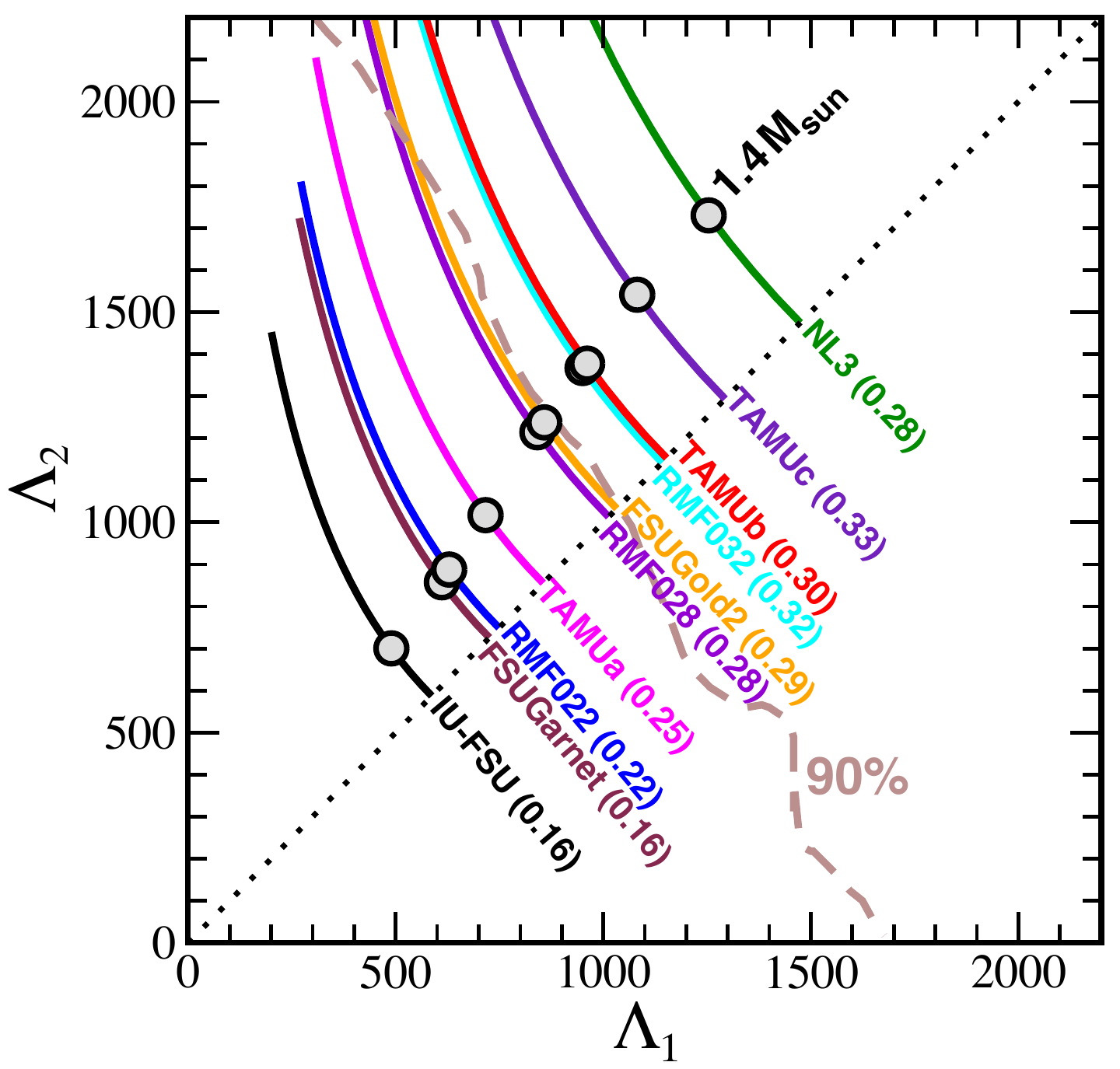}
 \caption{(Color online). Tidal polarizabilities $\Lambda_{1}$ and $\Lambda_{2}$
    associated with the high-mass $M_{1}$ and low-mass $M_{2}$
    components of the binary predicted by a set of
    ten distinct RMF models.}
  \label{Fig2}
\end{figure}
%%%%%%%%%%%%%%%%%%%%%%%%%%%%%%%%%%%%

In Fig.\,\ref{Fig2} we display predictions from all ten models for
the individual tidal polarizabilities $\Lambda_{1}$ and
$\Lambda_{2}$ associated with the high-mass $M_{1}$ and low-mass
$M_{2}$ components of the binary, respectively. The individual
curves are generated by allowing the high mass star to vary
independently within the $1.365\!\le M_{1}/M_{\odot}\!\le\!1.60$
range, whereas the low mass component is determined by maintaining
the chirp mass fixed at the observed value of
${\cal M}\!=\!(M_{1}M_{2})^{3/5}
(M_{1}\!+\!M_{2})^{-1/5}\!\!=\!1.188\,M_{\odot}$\,\cite{Abbott:PRL2017}.
Given that $R_{\rm skin}^{208}$ provides a proxy for the stiffness
of the symmetry energy near saturation density, we display in
parentheses the corresponding predictions for all ten models. Also
shown is the 90\% probability contour extracted from the low-spin
scenario assumed in Fig.\,5 of Ref.\,\cite{Abbott:PRL2017}.
%Note that the intersection of the model predictions with
%the $\Lambda_{1}\!=\!\Lambda_{2}$ line, corresponds to the equal mass
%case of $M_{1}\!=\!M_{2}\!=\!1.365\,M_{\odot}$.
For reference, we also highlight predictions for a binary system
having a high-mass component of $M_{1}\!=\!1.4\,M_{\odot}$
($M_{2}\!=\!1.33\,M_{\odot}$); this gives a rough indication of how
rapidly each model moves away from the equal-mass case
(denoted by the dotted line).
%scenario.
%Finally, itis important to underscore that the predictions displayed
%by each one of the models---which pertain to both finite nuclei and
%neutron stars---are generated from the same interaction; no
%parameters are allowed to change in going from finite nuclei to
%neutron stars.

As shown in Eq.\,(\ref{Lambda}), the tidal polarizability is highly
sensitive to the compactness of the neutron star. For a given mass,
models with a stiff symmetry energy (large $L$) are highly effective
in pushing against gravity, thereby generating large stellar radii
and correspondingly large tidal polarizabilities. The 90\% contour
recommended by the LIGO-Virgo collaboration is stringent enough
to disfavor overly stiff EOSs. Indeed,
%under the equal-mass scenario,
the four RMF models with the stiffest symmetry energy are ruled
out. The next two stiffest models considered here---FSUGold2 and
RMF028---follow closely the  90\% contour.

%However, as the mass asymmetry grows, the  90\% contour displays an
%intriguing behavior that is qualitatively different from all the
%model predictions (cf. Fig.\,5 of Ref.\,\cite{Abbott:PRL2017}.)
%That is, for $\Lambda_{1}\!\simeq\!500$, the 90\% contour indicates
%either a fairly rapid decrease in $\Lambda_{1}$ or a slower increase
%in $\Lambda_{2}$.In turn, this may be associated with either a rapid
%reduction of the stellar radius with increasing mass $M_{1}$ or a
%fairly constant value of the radius with decreasing mass $M_{2}$. If
%the former, this may be suggestive of a softening of the EOS at high
%density that could be associated with a phase transition. If the
%latter, this could indicate that the crust is relatively thin for
%low-mass stars corresponding to a small crust-core transition
%pressure\,\cite{Fattoyev:2010tb,Piekarewicz:2014lba}.

%%%%%%%%%%%%%%%%%%%%%%%%%%%%%%%%%%%%
%\vspace{1cm}
\begin{figure}[ht]
\smallskip
 \includegraphics[width=0.99\columnwidth]{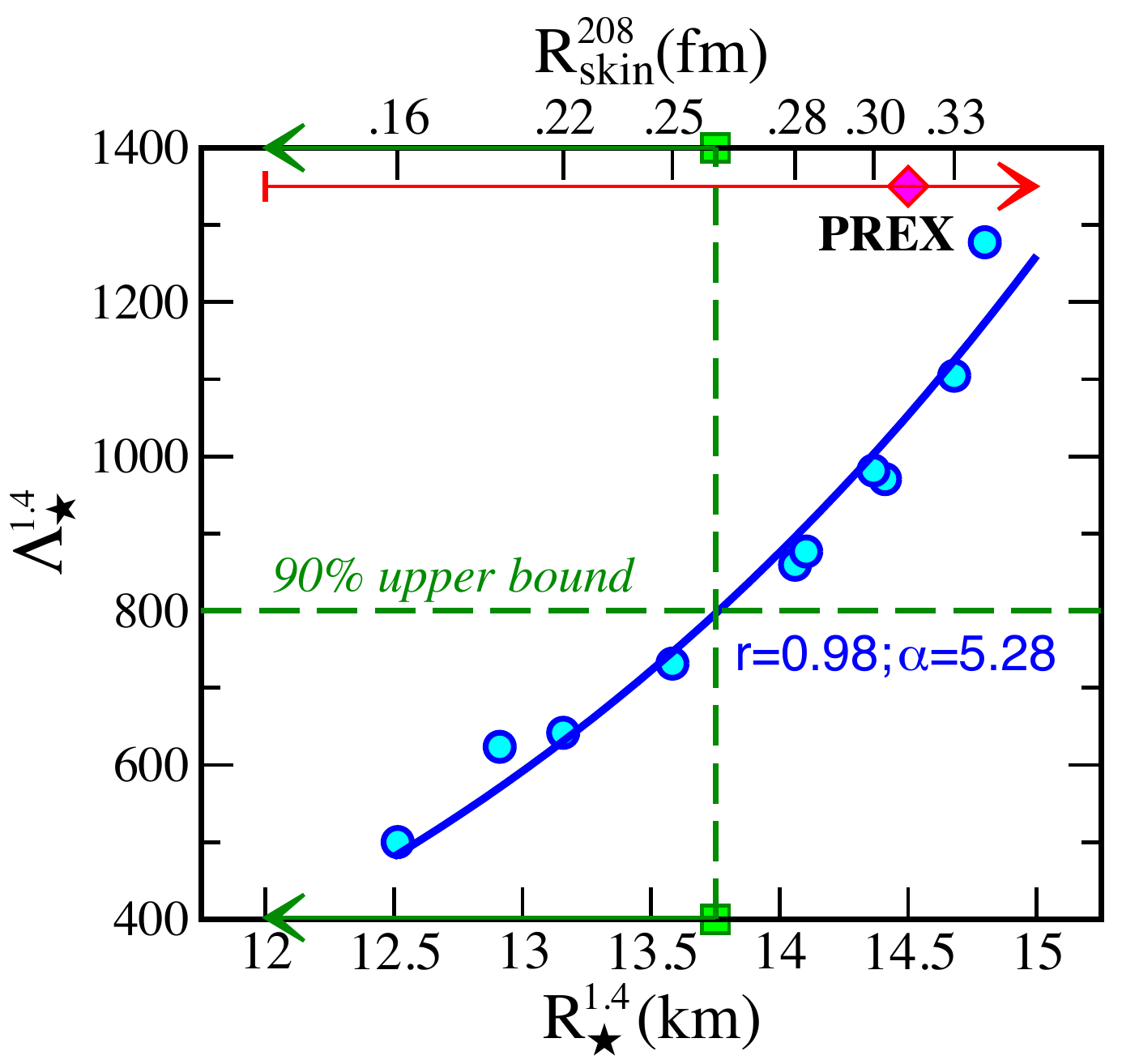}
 \caption{(Color online). As in Fig.\ref{Fig1}, predictions are shown
 for $\Lambda^{1.4}_{\star}$ as a function of the radius of a
 $1.4\,M_{\odot}$ neutron star and the neutron-skin thickness
 of $^{208}$Pb, but now for the ten RMF models discussed in the
 text.}
 \label{Fig3}
\end{figure}
%%%%%%%%%%%%%%%%%%%%%%%%%%%%%%%%%%%%

%However, as the mass asymmetry grows, the  90\% contour displays an
%intriguing behavior that is qualitatively different from all the
%model predictions (cf. Fig.\,5 of Ref.\,\cite{Abbott:PRL2017}.
%To explore this intriguing behavior further, we display)
In analogy to Fig.\,\ref{Fig1}, we display in Fig.\,\ref{Fig3} the tidal polarizability
of a $1.4\,M_{\odot}$ neutron star as a function of the corresponding stellar
radius and the neutron-skin thickness of $^{208}$Pb, but now for the ten
selected RMF models.
%: (a) the equal-mass
%case of $M_{1}\!=\!M_{2}\!=\!1.365\,M_{\odot}$ and (b) the most
%asymmetric case of $M_{1}\!=\!1.6\,M_{\odot}$ and
%$M_{1}\!=\!1.17\,M_{\odot}$, both fixed at a chirp mass value of
%${\cal M}\!=\!1.188\,M_{\odot}$.
%Upper limits for $\widetilde{\Lambda}$ are taken directly from
%Ref.\,\cite{Abbott:PRL2017}. For the equal-mass scenario, we plot
%%$\widetilde{\Lambda}$ against the predicted stellar radius of a
%$1.4\,M_{\odot}$ neutron star, while for the unequal-mass case
%against the radius of a $1.6\,M_{\odot}$ neutron star.
The solid line represents a two-parameter fit to the predictions of
the ten models of the form $\Lambda_{\star}\!=a R_{\star}^{\alpha}$.
We obtain $a\!\approx\!7.76\!\times\!10^{-4}$ and $\alpha\!\approx\!5.28$,
with a robust correlation coefficient of $r\!\approx\!0.98$. Note that the
exponent $\alpha$ is consistent with the scaling behavior suggested in
Eq.\,(\ref{Lambda}).
%$\Lambda_{\star}^{1.4}\!\approx\!\left(7.76\!\times\!10^{-4}\right)
%R_{\star}^{5.28}$ with a
%Given the strong correlation between the slope of the symmetry energy
%$L$ and the neutron-skin thickness of ${}^{208}$Pb, we also display model
%predictions for $R_{\rm skin}^{208}$.
Also note that predictions for tidal polarizabilities, stellar radii, and neutron
skins are made without ever changing the parameters of each individual model.

%%%%%%%%%%%%%%%%%%%%%%%%%%%%%%%%%%%%%%%%%%%%%%%%%%%%%%%%%%%%%%%%%%%%%
% FJF: In this paragraph I thought, it could maybe now better to use the
% FSUGold2 family instead, so updated this sentence below. Please
% revert back if you don't like it...

%Nevertheless, relying on the FSUGold2 model that
%predicts $R_{\star}^{1.4}\!<\!14.3\,{\rm km}$, one deduces the
%following limits: $R_{\rm skin}^{208}\!\lesssim\!0.29\,{\rm fm}$ and
%$L\!\lesssim\!112.7\,{\rm MeV}$.
%%%%%%%%%%%%%%%%%%%%%%%%%%%%%%%%%%%%%%%%%%%%%%%%%%%%%%%%%%%%%%%%%%%%%
As already alluded in Fig.\,\ref{Fig2}, limits imposed on the tidal
polarizability by GW170817 rule out the four models with the
stiffest symmetry energy. Now Fig.\,\ref{Fig3} illustrates how the
impact of the $\Lambda_{\star}^{1.4}\!\leq\!800$ limit translates
into a limit on the stellar radius of a $1.4 M_{\odot}$ neutron star
of $R_{\star}^{1.4}\!<\!13.76\,{\rm km}$. This is in excellent
agreement with the $R_{\star}^{1.4}\!<\!13.9\,{\rm km}$ limit
inferred previously from Fig.\,\ref{Fig1}. However, the
$\Lambda_{\star}^{1.4}\!\leq\!800$ limit is now stringent enough to
rule out all but the four models with the softest symmetry energy.
Given that both $L$ and $R_{\rm skin}^{208}$ are correlated to the
radius of ``low-mass'' neutron stars\,\cite{Carriere:2002bx},
deducing limits on these two quantities from the radius of a $1.4
M_{\odot}$ neutron star may be model dependent. Nevertheless, using
the stiffest of the models that survives the
$\Lambda_{\star}^{1.4}\!\leq\!800$ constraint as a guideline ({\sl
i.e.,} TAMUC-FSUa) one obtains: $R_{\star}^{1.4}\!=\!13.6\,{\rm
km}$, $R_{\rm skin}^{208}\!=\!0.25\,{\rm fm}$, and $L\!=\!82.5\,{\rm
MeV}$.

%Nevertheless, relying exclusively on the FSUGold2 model that
%predicts a value of $R_{\star}^{1.4}$ consistent with the above
%limit, one obtains:
%Note (FJF): I think this discussion may still be useful.
%{\color{red}{However, the situation changes dramatically under the
%asymmetric-mass scenario, when one adopts the limit of
%$\widetilde{\Lambda}\!\leq\!800$\,\cite{Abbott:PRL2017}. Using $M_1
%= 1.40 M_{\odot}$ and $M_2 = 1.33 M_{\odot}$, we find a fairly
%stringent limit on the radius of a $1.4\,M_{\odot}$ neutron star of
%$R_{\star}^{1.4}\!<\!13.36\,{\rm km}$. In this case only three of
%the original ten models---IU-FSU, FSUGarnet, and RMF022---remain
%consistent with the upper bound. By invoking a member of the
%FSUGold2 family that predicts a value of $R_{\star}^{1.4}$
%consistent with this limit, and assuming that the correlation
%between $R_{\star}^{1.4}$ and $R_{\rm skin}^{208}$ remains valid as
%one extrapolates down to nuclear densities, one infers the following
%limits: $R_{\rm skin}^{208}\!\lesssim\!0.24\,{\rm fm}$ and
%$L\!\lesssim\!77\,{\rm MeV}$. Besides the stringent constraints
%imposed on both $R_{\rm skin}^{208}$ and $L$, the alleged softening
%of the symmetry energy has profound consequences on many other
%phenomena, such as on the cooling of neutron stars and on the
%possible existence of exotic matter in the stellar core. Such
%interesting scenarios will be explored in a forthcoming paper.}}

%%%%%%%%%%%%%%%%%%%%%%%%%%%%%%%%%%%%%%%%%%%%%%%%%%%%%
%\vspace{1cm}
\begin{figure}[ht]
\smallskip
 \includegraphics[width=0.9\columnwidth,height=7.5cm]{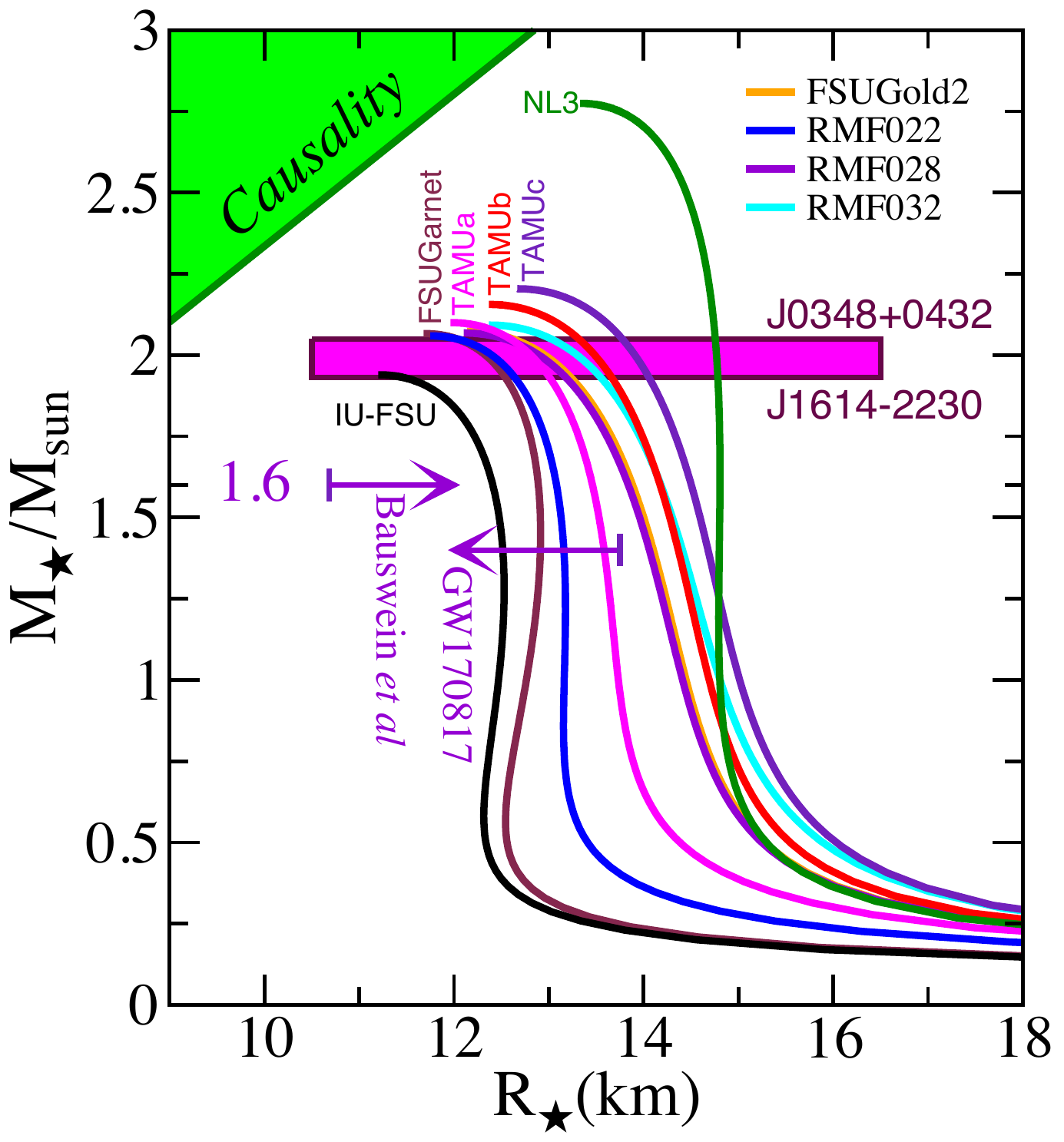}
 \caption{(Color online). Mass-vs-Radius relation predicted by the
 ten RMF models discussed in the text. Radius and mass constraints
 obtained from observation have been also incorporated into the plot.}
\label{Fig4}
\end{figure}
%%%%%%%%%%%%%%%%%%%%%%%%%%%%%%%%%%%%%%%%%%%%%%%%%%%%%

We conclude by displaying in Fig.\,\ref{Fig4} the ``holy-grail" of
neutron-star structure: the mass-{\sl vs}-radius (MR) relation. Note
that each EOS generates a unique MR relation. Interestingly, the
inverse statement is also true: knowledge of the MR relation
uniquely determines the EOS\,\cite{Lindblom:1992,Chen:2015zpa}.
Typically, the EOS is written as a sum of two distinct contributions:
(a) one for symmetric matter having equal number of neutrons and
protons and (b) one for the symmetry energy to account for deviations
from the symmetric limit. For RMF models of the kind described here,
the maximum stellar mass is largely controlled by the high-density
component of the EOS of symmetric matter. In contrast, stellar
radii---as well as tidal polarizabilities---are sensitive to the symmetry
energy at about twice nuclear-matter saturation density. However,
stellar radii are also sensitive to the EOS of the inhomogeneous
crust\,\cite{Piekarewicz:2014lba}. At densities relevant to the inner
crust, the system exhibits rich and complex structures that emerge
from a dynamical competition between short-range nuclear attraction
and long-range Coulomb repulsion. Due to this complexity, at present
the EOS of the inner crust is not well known. Hence, for this region we
have adopted the EOS described in Ref.\,\cite{Negele:1971vb}.
As already mentioned, all RMF models generate an EOS that is
sufficiently stiff to support a $M_{\star}\!\approx\!2\,M_{\odot}$
neutron star\,\cite{Demorest:2010bx,Antoniadis:2013pzd}.
%Note (FJF): Are we changing Fig. 4?
In addition, Fig.\,\ref{Fig4} incorporates our newly-inferred 13.76 km
upper limit on $R_{\star}^{1.4}$. Interestingly enough, a \emph{lower limit}
on the stellar radius of a $1.6\,M_{\odot}$ neutron star of
$R_{\star}^{1.6}\!=\!10.68_{-0.04}^{+0.15}$ was obtained by
Bauswein {\sl et al.,} under the assumption that the BNS merger
did not result in a prompt collapse\,\cite{Bauswein:2017vtn}.
%By combining this result with the nearly perfect fit displayed on
%the right-hand panel of Fig.\,\ref{Fig3}, one can also set a lower
%limit on the tidal polarizability; that is,
%$\!250\!\leq\!\widetilde{\Lambda}\!\leq\!800$. A further
%extrapolation suggests that $\widetilde{\Lambda}\! \simeq \!250$
%corresponds to $R_{\rm skin}^{208}\!\simeq\!0.04$ fm.
Finally, we use the results obtained in Fig.\,\ref{Fig3}
to deduce a \emph{lower limit} on the tidal polarizability of a
$1.4\,M_{\odot}$ neutron star. To do so, we note that PREX
imposes a lower bound on the neutron-skin thickness of
${}^{208}$Pb of $R_{\rm skin}^{208}\!\simeq\!0.15\,{\rm fm}$,
which corresponds to a stellar radius of
$R_{\star}^{1.4}\!\simeq\!12.55\,{\rm km}$. Using the fit
displayed in Fig.\,\ref{Fig3}, the limit on $R_{\star}^{1.4}$ translates
into a corresponding lower limit on the tidal polarizability of
$\Lambda_{\star}^{1.4}\!\simeq \!490$; see Ref.\,\cite{Radice:2017lry}
for an alternative extraction of a lower bound on the tidal deformability
parameter. Thus, combining observational constraints from the LIGO-Virgo
collaboration with laboratory constraints from the PREX collaboration,
the tidal polarizability of a $1.4\,M_{\odot}$ neutron star falls within
the following range of values:
$490\!\lesssim\!\Lambda_{\star}^{1.4}\!\lesssim\!800$.

%This suggests that $R_{\rm skin}^{208}$ and $R_{\star}^{1.6}$ probe
%different regions of the symmetry energy\,\cite{Fattoyev:2014pja}.

%A further extrapolation suggests that the lower value on $R_{\rm
%skin}^{208}$ can be negative. While the negative value of the skin
%thickness contradicts the current PREX results, this also indicates
%that extrapolating from densities typical of a $1.6\,M_{\odot}$
%neutron star to the laboratory can be risky.

%%%%%%%%%%%%%%%%%%%%%%%%%%%%%%%%%%%%%%%%%%%%%%%%%%%%%%%%%%%%%%%%%%%%
% FJF: I think, we could shorten the following paragraph
% but I am leaving it as long at the moment. We may need to cut some
% spaces to make it four pages.
%%%%%%%%%%%%%%%%%%%%%%%%%%%%%%%%%%%%%%%%%%%%%%%%%%%%%%%%%%%%%%%%%%%%

%The determination of stellar radii by photometric means has
%been plagued by large systematic uncertainties, often revealing discrepancies as large as
%5-6\,km\,\cite{Ozel:2010fw,Steiner:2010fz,Suleimanov:2010th}. It
%appears, however, that the situation has improved through a better
%understanding of systematic uncertainties, important theoretical
%developments, and the implementation of robust statistical
%methods\,\cite{Guillot:2013wu,
%Lattimer:2013hma,Heinke:2014xaa,Guillot:2014lla,Ozel:2015fia,Watts:2016uzu,
%Steiner:2017vmg,Nattila:2017wtj}. Nevertheless, that a single
%detection of a BNS merger already places important constraints on
%stellar radii,  %{\sl i.e.,} $R_{\star}^{1.6}\!=\!(10.68$--$13.25)\,{\rm km}$
%and on the neutron thickness of $^{208}$Pb, is truly a remarkable
%achievement.

In summary, we have examined how the historical first detection of
gravitational waves from the merger of two neutron stars improves
our knowledge of the EOS of dense matter. While the BNS merger
provides fundamental insights on the site of the $r$-process and
confirms its association to short $\gamma$-ray burst, our aim in
this letter was to illuminate its connection to laboratory
observables. Such a connection is possible because of the
sensitivity of the tidal polarizability to the stellar radius, which probes
the symmetry energy at about twice nuclear-matter saturation density.
Assuming that one can extrapolate down to saturation density, constraints
from GW170817 provide limits on the neutron-skin thickness of
${}^{208}$Pb---a fundamental laboratory observable that is strongly
correlated to the slope of the symmetry energy at saturation density.
Indeed, by exploring the consequences of the
$\Lambda_{\star}^{1.4}\!\leq\!800$ limit provided by the LIGO-Virgo
collaboration, we deduced a limit on the stellar radius of a
$1.4\,M_{\odot}$ neutron star of $R_{\star}^{1.4}\!<\!13.76\,{\rm km}$.
In turn, this translates into a neutron-skin thickness of ${}^{208}$Pb of
$R_{\rm skin}^{208}\!\lesssim\!0.25\,{\rm fm}$, which is well below
the upper limit obtained by the PREX collaboration. Conversely, by
relying on PREX lower limit on $R_{\rm skin}^{208}$, we were able
to provide a lower limit on the tidal polarizability of
$\Lambda_{\star}^{1.4}\!\gtrsim\!490$. Finally, given that the PREX
experiment reported a central value of
$R_{\rm skin}^{208}\!\lesssim\!0.33\,{\rm fm}$---albeit with large error
bars---an intriguing possibility emerges. If the follow-up experiment
PREX-II confirms that $R_{\rm skin}^{208}$ is large, this will suggest
that the EOS at the typical densities found in atomic nuclei is stiff. In
contrast, the relatively small neutron-star radii suggested by GW170817
implies that the symmetry energy at higher densities is soft. The
evolution from stiff to soft may be indicative of a phase transition
in the neutron-star interior. Undoubtedly, the multi-messenger era
is in its infancy and much work remains to be done. Yet, it is remarkable
that the very first observation of a BNS merger already provides a
treasure trove of insights into the nature of dense matter.

%%%%%%%%%%%%%%%%%%%%%%%%%%%%%%%%%%%%%%%%%%%%%%%%%%%%%%%%%%%%%%%%%

%\vspace{-0.6cm}
\begin{acknowledgments}
 We are grateful to Katerina Chatziioannou and Jocelyn Read for
 clarifying the LIGO-Virgo results presented in
 Ref.\,\cite{Abbott:PRL2017}.
 This material is based upon work supported by the U.S. Department
 of Energy Office of Science, Office of Nuclear Physics under Awards
 DE-FG02-87ER40365 (Indiana University), Number DE-FG02-92ER40750 (Florida
 State University), and Number DE-SC0008808 (NUCLEI SciDAC Collaboration).
\end{acknowledgments}

\bibliography{ReferencesJP}
%\bibliography{GW170817.bbl}

%\vfill\eject
\end{document}